\documentclass[journal=jpclcd,manuscript=letter,linenumber]{achemso}
\usepackage{color}
\usepackage{soul}
\sethlcolor{yellow}
\usepackage{tabularx}
\usepackage{graphicx}
\usepackage{bm}
\usepackage{float}
\usepackage[version=3]{mhchem}

\newcommand{\ket}[1]{\left| #1 \right\rangle}
\newcommand{\bra}[1]{\left\langle #1 \right|}

\newcommand{\bv}[1]{\mathbf{#1}}

\newcommand{\COMMENT}[1]{}
\newcommand{\textapprox}{{\raise.17ex\hbox{$\scriptstyle\mathtt{\sim}$}}}

 \author{Fan Zheng}
 \affiliation{The Makineni Theoretical Laboratories, Department of Chemistry, University of Pennsylvania, Philadelphia, PA 19104--6323 USA}
 \author{Hiroyuki Takenaka}
 \affiliation{The Makineni Theoretical Laboratories, Department of Chemistry, University of Pennsylvania, Philadelphia, PA 19104--6323 USA}
 \author{Fenggong Wang}
 \affiliation{The Makineni Theoretical Laboratories, Department of Chemistry, University of Pennsylvania, Philadelphia, PA 19104--6323 USA} 
 \author{Nathan Z. Koocher}
 \affiliation{The Makineni Theoretical Laboratories, Department of Chemistry, University of Pennsylvania, Philadelphia, PA 19104--6323 USA}
 \author{Andrew M. Rappe}
 \email{rappe@sas.upenn.edu}
 \affiliation{The Makineni Theoretical Laboratories, Department of Chemistry, University of Pennsylvania, Philadelphia, PA 19104--6323 USA}

 \title{First-principles calculation of bulk photovoltaic effect in CH$_3$NH$_3$PbI$_3$ and CH$_3$NH$_3$PbI$_{3-x}$Cl$_{x}$}


\begin{document}
  
 \begin{abstract}
  Hybrid halide perovskites exhibit nearly 20\% power conversion efficiency, but the origin of their high efficiency is still unknown. Here, we compute the shift current, a dominant mechanism of bulk photovoltaic (PV) effect for ferroelectric photovoltaics, in CH$_3$NH$_3$PbI$_3$ and CH$_3$NH$_3$PbI$_{3-x}$Cl$_{x}$ from first principles. We find that these materials give approximately three times larger shift current PV response to near-IR and visible light than the prototypical ferroelectric photovoltaic BiFeO$_3$. The molecular orientations of CH$_3$NH$_3^{+}$ can strongly affect the corresponding PbI$_3$ inorganic frame so as to alter the magnitude of the shift current response.  Specifically, configurations with dipole moments aligned in parallel distort the inorganic PbI$_3$ frame more significantly than configurations with near net zero dipole, yielding a larger shift current response. Furthermore, we explore the effect of Cl substitution on shift current, and find that Cl substitution at the equatorial site induces a larger response than does substitution at the apical site.  
 \end{abstract}



  The power conversion efficiency of hybrid halide perovskites has almost doubled from 9.7\%~\cite{Kim12p591} to 19.3\%~\cite{Zhou14p542} within two years. This attracts a great deal of interest in understanding the mechanism of its photovoltaic (PV) effect and in designing and optimizing the materials. Different synthetic methods and cell architectures can affect the final efficiency, implying that various PV mechanisms may contribute to the PV current~\cite{Gao14px}. Particularly, the doping of Cl into CH$_3$NH$_3$PbI$_3$ (methylammonium lead iodide, MAPbI) increases the diffusion length of excited carriers by nearly one order of magnitude~\cite{Stranks13p341}. Significant current effort focuses on understanding the underlying reason for this high PV efficiency. Both experimental measurements and theoretical calculations show that MAPbI and related materials, such as NH$_2$CHNH$_2$PbI$_3$ (formamidinium lead iodide, FAPbI), MAPbI$_{3-x}$Cl$_{x}$ (MAPbICl), MAPb$_x$Sn$_{1-x}$I (MAPbSnI) and MAPbI$_{3-x}$Br$_{x}$ (MAPbIBr), have band gaps in the range of 1.1-2.1 eV, in the visible light region~\cite{Papavassiliou95p1713,Noel14px,Umari14p4467,Chiarella08p045129,Ogomi14p1004,Eperon14p982, Stoumpos13p9019,Mosconi14p16137}. Their optical absorption strength is comparable to other classic semiconductors such as GaAs, InP and CdTe~\cite{Green14p506,Feng14p1278,Filippetti14p125203,Umari14p4467}. These materials also show relatively high and balanced electron and hole mobility and very fast electron-hole pair generation~\cite{Ponseca14p5189,Xing13p344,Edri14p3461}. In addition, the carrier diffusion length is more than 1 $\mu$m, implying a low concentration of deep defects~\cite{Stranks13p341,Kim14p1312,Yin14p063903,Du14p9091}. The source of their high power conversion efficiency, however, is still not clear.
  
  Shift current has been proven to be a main mechanism of the bulk photovoltaic effect (BPVE) in ferroelectric oxides such as BaTiO$_3$, PbTiO$_3$ and BiFeO$_3$~\cite{Young12p116601,Young12p236601,Sipe00p5337,vonBaltz81p5590}. Single-phase noncentrosymmetric materials are able to generate DC current under uniform illumination. Shift current relies on ballistic quantum coherent carriers, so it can provide above band gap open-circuit photo-voltage. MAPbI and related materials share similar perovskite structures with ferroelectric oxides. The tetragonal phase of MAPbI was found to have ferroelectric response at room temperature~\cite{Stoumpos13p9019}. Various $I/V$ hysteresis measurements suggest that the current is related to the ferroelectric response~\cite{Snaith14p1511,Sanchez14p2357,Dualeh13p362,Docampo14p081508,Jeon14p1}. In particular, the large measured open-circuit voltage allows for the possibility that the BPVE could make a big contribution to the photo-voltage, as the BPVE can generate a photo-voltage that is above a material's band gap. Therefore, studying the BPVE of MAPbI-based materials is important in terms of elucidating the underlying mechanism of their high efficiency and continuously optimizing their properties as a solar cell material. In this letter, we calculate the shift current response of MAPbI and MAPbICl, and show that their current responses are approximately three times larger than that of BiFeO$_3$. Our calculations demonstrate that the molecular orientations as well as the Cl substitution position can strongly affect their shift current responses.
  
  As shown in Ref.~\cite{Young12p116601}, the short-circuit shift current response $\sigma$ is a rank three tensor, and it can be computed using perturbation theory, yielding the formula in the thin sample limit as
  \begin{flalign}\label{eq:sc_equ}
   J_q & = \sigma_{rsq}E_rE_s  \nonumber\\
   \sigma_{rsq}(\omega) & = \pi e \left(\frac{e}{m\hbar \omega}\right)^2 \sum_{n',n''} \int d\bv{k} \left(f[n''\bv{k}]-f[n'\bv{k}]\right)  \nonumber \\
                      & \quad \times \bra{n'\bv{k}}\hat{P}_r\ket{n''\bv{k}}\bra{n''\bv{k}}\hat{P}_s\ket{n'\bv{k}} \nonumber \\
                      & \quad \times \left(-\frac{\partial \phi_{n'n''}(\bv{k},\bv{k})}{\partial k_q}-\left[\chi_{n''q}(\bv{k})-\chi_{n'q}(\bv{k})\right]\right) \nonumber \\
                      & \quad \times \delta \left(\omega_{n''}(\bv{k})-\omega_{n'}(\bv{k})\pm \omega\right)
  \end{flalign}
  where $n$ and $\bv{k}$ are, respectively, the band index and wave-vector, $f$ is the occupation, $\hbar\omega_n$ is the energy of state $n$, $\phi_{n',n''}$ is the phase of the momentum matrix element between state $n'$ and $n''$ and $\chi_n$ is the Berry connection for state $n$. If spin-orbit coupling is considered, each Bloch state has spinor form and the current response becomes a sum over spinor components. In a thick sample, considering the light absorption coefficient $\alpha_{rr}\left(\omega\right)$, the current response can be described by the Glass coefficient $G$~\cite{Glass74p233}
  \begin{flalign}
   G_{rrq}=\frac{\sigma_{rrq}}{\alpha_{rr}}
  \end{flalign}
  When measuring in-plane current, the total current $\bv{J}$ is $J_q(\omega) = G_{rrq}I_r w$, where $I$ is the light intensity and $w$ is the sample width.
  
  \begin{figure}[h]
      \includegraphics[width=4.5in]{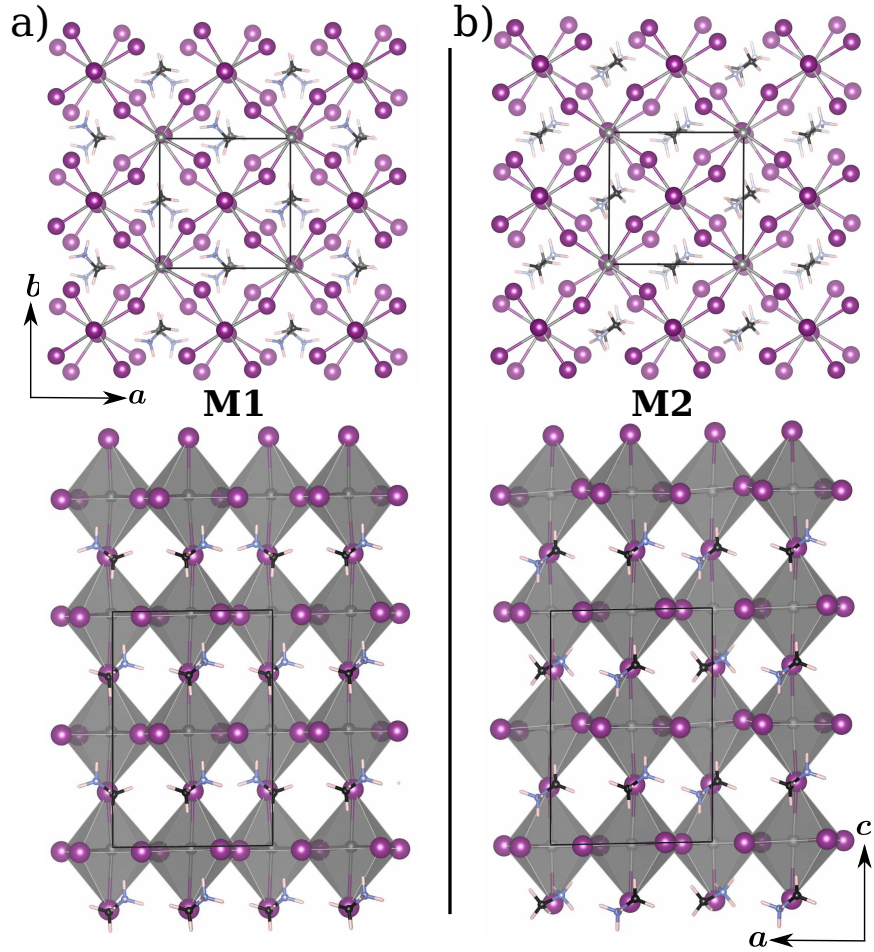}
      \caption{The top and side views of the relaxed MAPbI structures with a) molecular orientation 1 (M1) and b) molecular orientation 2 (M2). M1 has all the net MA molecular dipoles along the $c$ axis, while M2 has MA molecules with dipoles opposite to that of its neighboring molecules, yielding a net zero dipole. Four MAPbI$_3$ are considered in one unit cell. Pb: dark grey, I: purple, C: black, N: light blue, Cl: green, H: light pink.}\label{two_molecule_ab}
  \end{figure}

  The plane-wave density functional theory (DFT) package QUANTUM-ESPRESSO was used with generalized gradient approximation functional (GGA) to perform structure relaxations and electronic structure calculations ~\cite{Giannozzi09p395502}. Norm-conserving, designed nonlocal pseudopotentials were generated with the OPIUM package~\cite{Rappe90p1227, Ramer99p12471}. A plane-wave cutoff energy of 50 Ry was sufficient to converge the total energy. All the structures were fully relaxed with a force tolerance of 0.005 eV/\AA. In order to compute the shift current and Glass coefficient response tensors as shown in Eq.1, a self-consistent calculation with a $6\times6\times4$ $k$-point grid was performed to find the converged charge density. Then, a non self-consistent calculation with a much denser $k$-point grid was performed as needed to converge the shift current response. Due to the presence of heavy atoms, spin-orbit coupling is thought to affect the electronic structure and, by extension, the shift current~\cite{Amat14p3608, Even13p2999}. When spin-orbit coupling is included, an angular momentum dependent term is added to the Hamiltonian.  In this work, Hamiltonians without spin-orbit coupling (NSOC) and with spin-orbit coupling (SOC) are considered for electronic structure and shift current calculations, as prediction of a correct band gap is important for shift current calculations.

  MAPbI has perovskite-type structure with methylammonium (MA) at the $A$ site and Pb at the $B$ site. The orthorhombic structure is favored at low temperature, but with increasing temperature, it transforms to the tetragonal phase with the $I4/mcm$ space group~\cite{Poglitsch87p6373,Stoumpos13p9019}. The increased symmetry at high temperature is related to the free rotation of the MA molecules, as observed in both experiments and theoretical calculations~\cite{Poglitsch87p6373, Quarti13p279,Frost14p2584, Brivio13p042111, Mosconi14p16137,Wang14p1424}. In order to explore the effect of molecular orientation on the structure and shift current, structures with two different orientations, M1 and M2, starting from the tetragonal PbI$_3$ inorganic frame, are computed and shown in Fig.~\ref{two_molecule_ab}.  The lattice constants of the relaxed structures, as well as the experimental lattice constants, are shown in Table~\ref{table_lattice}.
  \begin{table*}[h]
  \caption{The lattice constants and relative total energies, per unit cell, of the optimized MAPbI structures with molecular orientation M1 and molecular orientation M2. The experimental values are from 
  Ref.~\cite{Stoumpos13p9019,Poglitsch87p6373,Baikie13p5628}. Total energy (per 48 atom cell) of the M2 orientation structure is set to zero.}\label{table_lattice}
  \begin{tabular}{ccccccc}\hline\hline
                   &  {\ } &  {M1}  &  {\ }  & {M2} & {\ } & exp. \cr\hline
          $a$ (\AA) &  {\ } & 8.97   &  {\ }  & 9.00 & {\ } & 8.85 \cr 
          $b$ (\AA) &  {\ } & 8.86   &  {\ }  & 8.77 & {\ } & 8.85  \cr
          $c$ (\AA) &  {\ } & 12.85  &  {\ }  & 12.95& {\ } &12.44--12.66  \cr
  $E_{\rm{T}}$ (eV) &  {\ } & 0.021     &  {\ }  & 0 & {\ } &  - \cr \hline \hline
  \end{tabular}
  \end{table*}
  As shown in the table, our calculated lattice constants agree well with the experiments and theoretical works~\cite{Kim14p1312,Umari14p4467,Amat14p3608,Mosconi13p13902,Wang14p1424,Egger14p2728}. The computed $a$ and $b$ lattice constants are slightly different depending on the molecular orientations, and because they are different, this confirms that the orthorhombic structure is favored at low temperature. Although the molecular orientations affect the lattice, the total energy difference between the M1 and M2 structures is small.

  \begin{figure}[h]
      \includegraphics[width=4.5in]{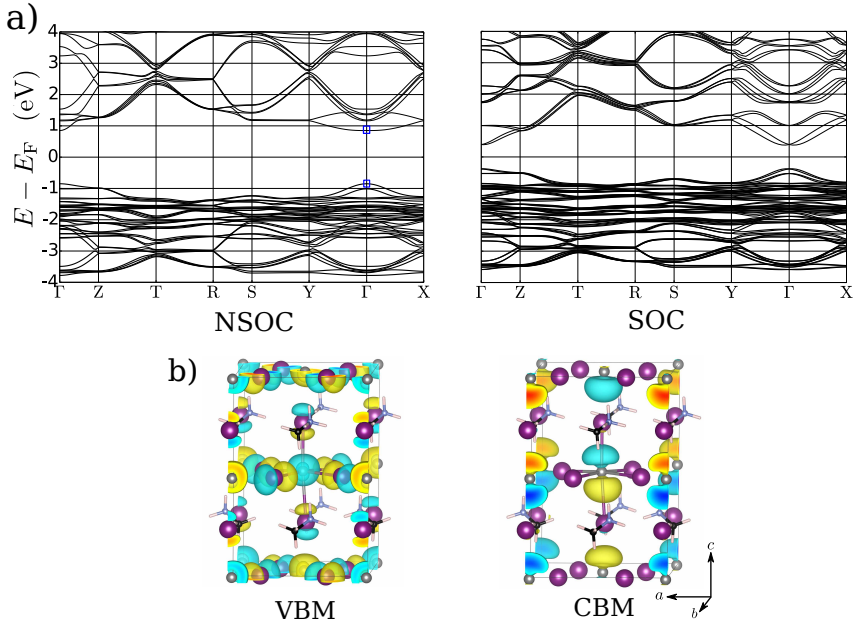}
      \caption{a) The band structures of MAPbI (M1) without and with spin-orbit coupling. Since the system lacks inversion symmetry, the SOC splits bands which are originally degenerate without SOC. b) The wavefunctions of the VBM and CBM at the $\Gamma$ point without SOC (VBM and CBM states are indicated as blue square in the NSOC band structure).}\label{bands_}
  \end{figure}

  Fig.~\ref{bands_} a) shows the band structures without and with spin-orbit coupling. As seen in the figure, SOC reduces the band gap substantially at the $\Gamma$ point. At this point, the GGA band gap with NSOC is close to the experimental value, as the DFT underestimation of band gap is largely canceled by the exclusion of SOC.  This has been seen in $GW$ and hybrid functional calculations~\cite{Umari14p4467,Feng14p1278}.  Our calculated electronic structure (NSOC) shows that the conduction band minimum (CBM) has mostly non-bonding Pb $p_z$ orbital character slightly hybridized with I $s$; whereas the valence band maximum (VBM) is anti-bonding between I $p$ and Pb $s$ orbitals, as shown by the real space wavefunction plots in Fig.~\ref{bands_} b). The MA molecular electronic states are not directly involved in the states near the band gap, as confirmed by other first-principles calculations~\cite{Frost14p2584,Filippetti14p125203}. 
  
\begin{figure}[h]
     \includegraphics[width=5.4in]{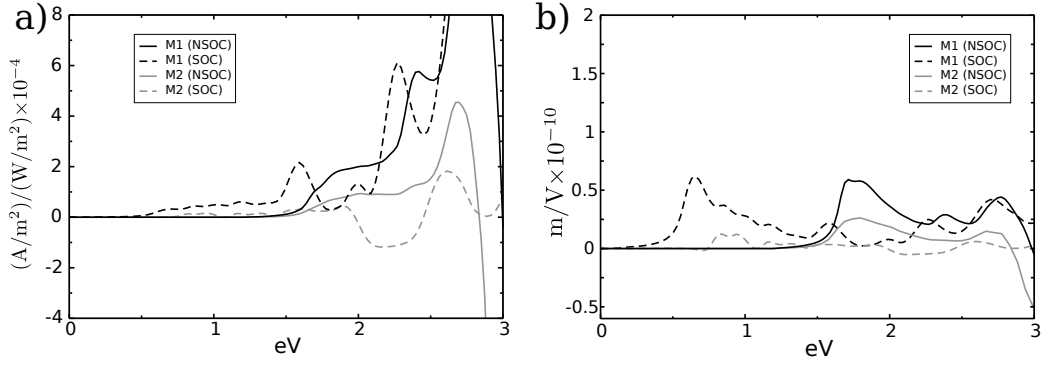}
     \caption{a) Shift current response $\sigma_{zzZ}$ and b) Glass coefficient $G_{zzZ}$ vs. incident photon energy for structures with molecular orientation M1 and orientation M2. The M1 structure provides a larger shift current response and Glass coefficient than the M2 structure. Calculations with and without SOC show the same trend for the two orientations. } \label{m_}
  \end{figure}

  Because the $zzZ$ response tensor component is the dominant component among all the tensor elements, Fig.~\ref{m_} shows the MAPbI thin sample limit shift current response $\sigma_{zzZ}$ and Glass coefficient response $G_{zzZ}$ for the M1 and M2 structures with and without SOC.  As seen in the figure, the response spectrum with SOC has a smaller band gap compared to the NSOC spectrum. Also, the SOC tends to shift the whole spectrum without substantially changing its magnitude. On average, the M2 structure has a much smaller current response and Glass coefficient than the M1 structure. The magnitude of the Glass coefficient is closely related to material symmetry and state delocalization~\cite{Young12p116601}. We have shown that a strongly distorted structure with delocalized states involved in an optical transition tends to give a large Glass coefficient response. Polarization calculations show that the M1 structure has a polarization of 5$\rm{\mu C/cm^2}$, while the M2 structure has nearly zero polarization. Since the bulk polarization contribution from the molecular dipole moment is estimated to be less than 2.5$\rm{\mu C/cm^2}$~\cite{dipolemoment,Dykstra11p1167,Schmidt93p1347}, the PbI$_3$ inorganic lattice is a significant contributor to the M1 structure's polarization, as much larger Pb displacement ($\approx$ 0.07 {\AA} along $z$) was observed than in M2 structure ($\approx$ 0.01 {\AA} along $z$). As a result, the distorted M1 structure provides a larger shift current response than the more symmetric M2 structure. At room temperature, the shift current responses can be the average of the M1 and M2 cases due to the disordered molecular orientations. Limiting the molecular rotation by methods such as doping, lattice shrinkage, or application of electric field can highlight the current contribution from one particular orientation. An understanding of the dependence of the current on molecular and polarization orientations is helpful in understanding the $I/V$ hysteresis under different applied voltage scanning rates.
  
\begin{figure}[h]
      \includegraphics[width=2.5in]{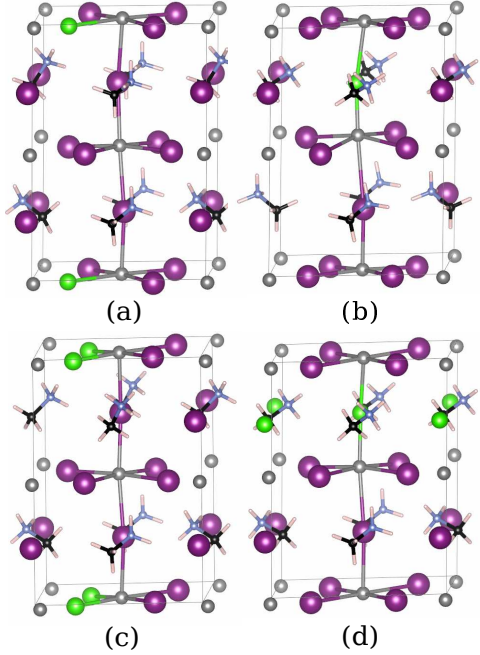}
      \caption{Fully relaxed structures of MAPbICl with one (a and b) or two (c and d) Cl atoms per unit cell. The structures shown here have molecular orientation M1. Structures a and c have equatorial site substitution; structures b and d have apical substitution. The corresponding four structures with molecular orientation M2 are also tested, but are not shown here. }\label{cl_}
  \end{figure}

  Cl doping helps to stabilize the crystallization of MAPbI by introducing lattice distortion~\cite{Gao14px}. Furthermore, adding Cl has been shown to provide a diffusion length as long as 1 $\mu$m without substantially changing the absorption spectrum~\cite{Stranks13p341,Filippetti14p125203,Green14p506}. Since MAPbICl has been found to have a reduced lattice constant along the $c$ axis compared to MAPbI, it is thought that Cl substitutes I at the apical site of the PbI$_6$ octahedra, but the actual Cl position is still not clear~\cite{Lee12p643,Colella14px}. In order to understand the effects of Cl position on shift current, we study different Cl substitution configurations at the apical and equatorial sites for both molecular orientations, as shown in Fig.~\ref{cl_}.
  
  The structures with one and two Cl atoms in one unit cell (with 48 atoms) are fully relaxed, and their band gaps and polarization magnitudes are shown in Table~\ref{table_}. Analysis of the total energies reveals that the structures with Cl at apical sites are slightly more favorable than the structures with Cl at equatorial sites.  This is because the equatorial substitutions introduce larger distortions into the octahedra. We find that although the Cl position has no substantial effect on the polarization, it strongly affects the shift current responses.
  \begin{figure}[h]
      \includegraphics[width=5in]{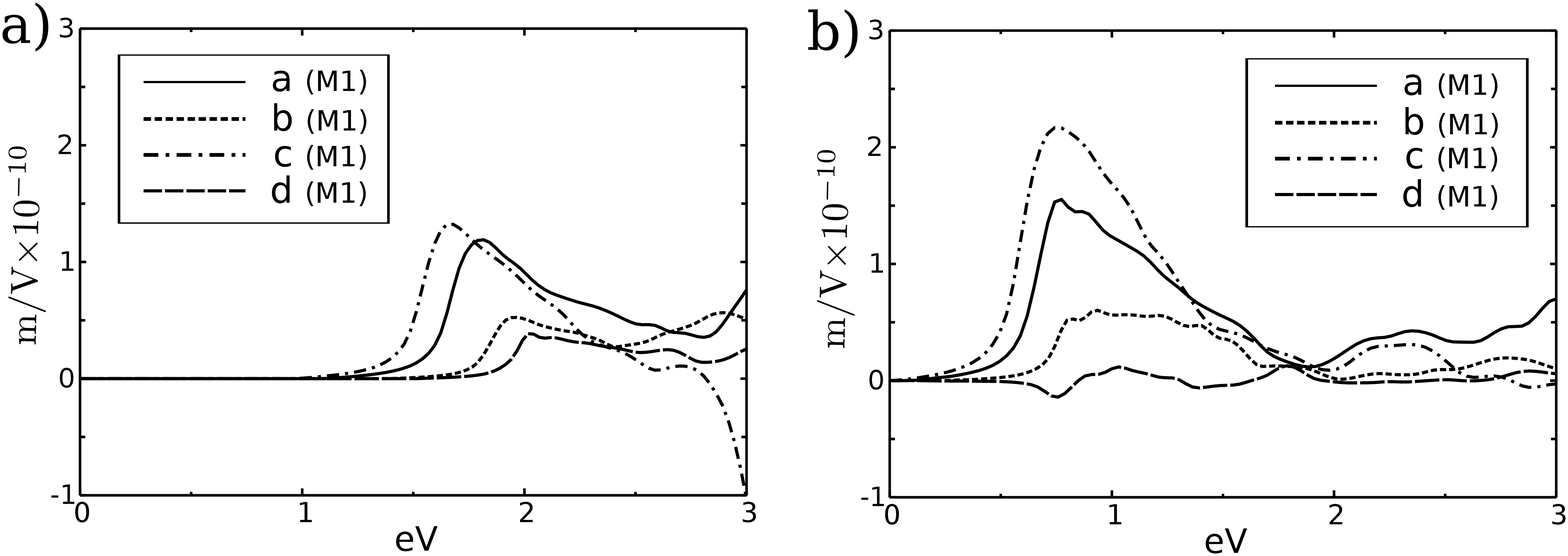}
      \caption{Glass coefficient $G_{zzZ}$ calculated with a) NSOC and b) SOC vs. incident photon energy for the four relaxed structures (a, b, c and d) shown in} Fig.~\ref{cl_}.\label{m1_cl_}
  \end{figure} 
  Fig.~\ref{m1_cl_} shows the Glass coefficient $G_{zzZ}$. Responses with and without SOC show a similar trend for different Cl positions and concentrations. Interestingly, the apical site substitution of I with Cl tends to give relatively small shift current responses, while the equatorial site substitution shows much larger responses. This can be explained from wavefunction projections. 
  \begin{figure}[h]
      \includegraphics[width=4.in]{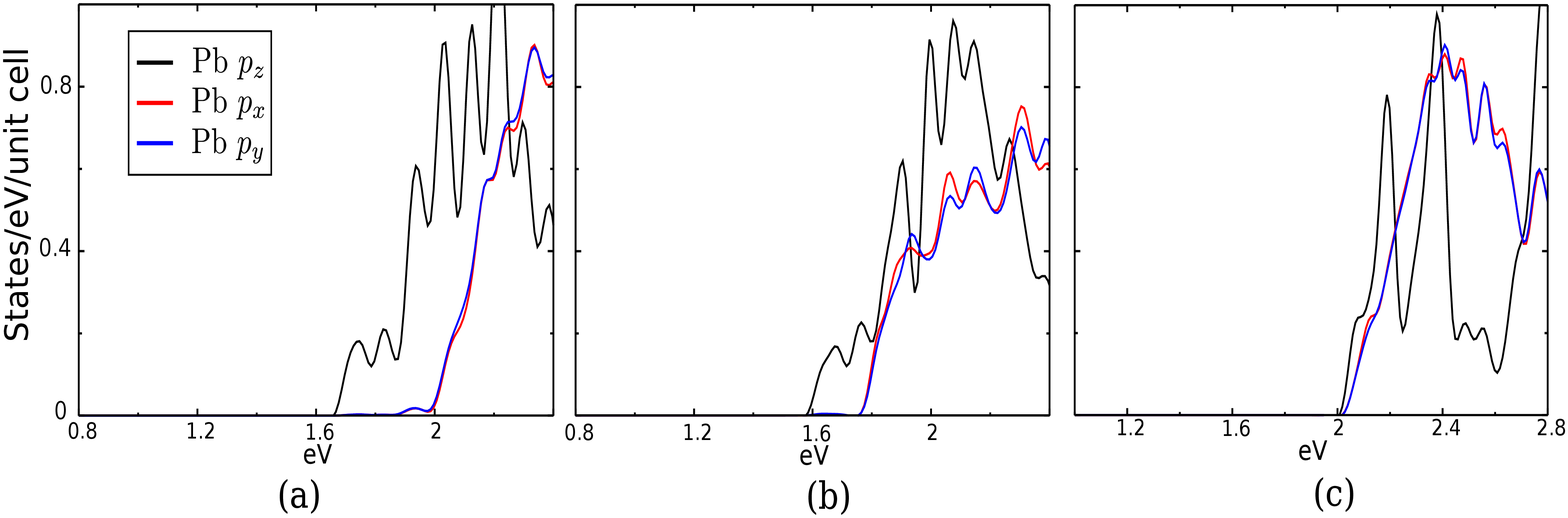}
      \caption{The projected density of states (PDOS) for Pb atoms calculated from structure a) M1 orientation without Cl in Fig.~\ref{two_molecule_ab}, b) structure c(M1) and c) d(M1) in Fig.~\ref{cl_}. Energy of the VBM is set to zero.}\label{pb_pdos_}
  \end{figure}
  Our electronic structure calculations show that Pb {\em p} orbital character slightly hybridized with I dominates the conduction bands near the band gap. In the highly symmetric structure without octahedral tilting, Pb $p_x$, $p_y$ and $p_z$ will be degenerate and hybridized with I $s$. However, the distortion of Pb-I bonds on the $a$-$b$ plane will cause the Pb $p$ orbitals to hybridize with I $p$ orbitals in addition to I $s$ orbitals. This will lift the original degeneracy between $p_{x/y}$ and $p_z$, allowing Pb $p_z$ to become the dominant orbital character of the CBM. This is very clear in the NSOC case. Wave functions calculated with SOC show a similar picture, but it is not as obvious as in the NSOC case since orbitals with different angular momentum are mixed together. We can see from the projected density of states (PDOS) calculated without SOC (Fig.~\ref{pb_pdos_}) that for the structure without Cl, the CBM has mostly Pb $p_z$ orbital character. With increasing Cl concentration at the apical site, the strong electronegativity of Cl increases the energy bands with Pb $p_z$ orbital character while leaving $p_{x/y}$ unchanged, allowing for a larger band gap. The CBM state, mainly composed of hybridized $p_{x/y}$ orbital character, reduces the current flowing along $z$ direction, since the $p_{x/y}$ orbitals are less delocalized than $p_z$ orbitals along $z$.  However, the Cl concentration only moderately affects the shift current response, as structures containing one Cl atom yield responses similar in magnitude to those of structures containing two Cl atoms.

  \begin{figure}[h]
      \includegraphics[width=5in]{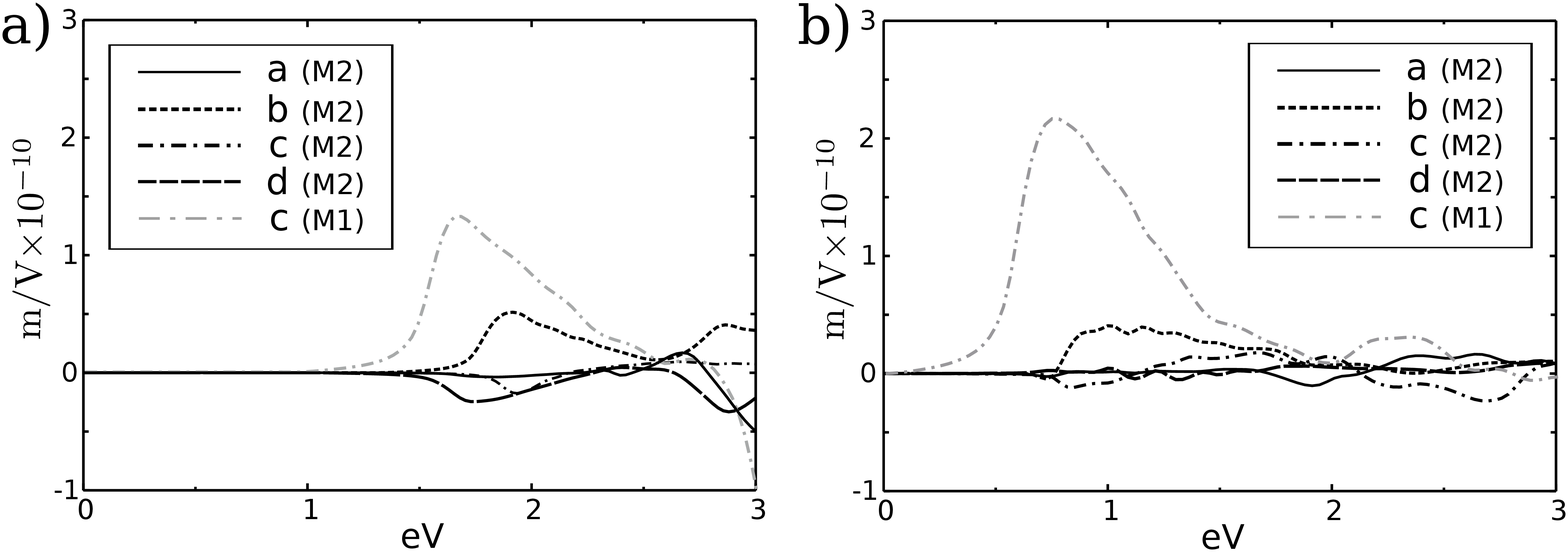}
      \caption{Glass coefficient $G_{zzZ}$ a) without and b) with SOC for the four relaxed structures (a, b, c and d) shown in Fig.~\ref{cl_}, but with molecular orientation M2. For comparison, the largest response from the M1 case (structure c(M1)) is also plotted as a grey line. On average, the M2 orientation gives smaller responses than the M1 orientation for the four structures.}\label{m2_cl_}
  \end{figure} 
  
  The shift current is also calculated for the molecular orientation M2, which has the same Cl configurations discussed previously.  Overall, their responses, shown in Fig.~\ref{m2_cl_}, are smaller than that corresponding to the M1 orientation, and are similar to the case without Cl. The minor electronic contribution of the organic species at the band edge, evident from the PDOS, suggests that the effect of the molecular orientation on the shift current is likely indirect, occurring through the PbI$_{3}$ frame. In this case, because the molecular dipoles for the neighboring molecules are opposite, there is no net dipolar effect on the PbI$_3$ frame, resulting in a nearly symmetric frame. As the distortion decreases, the dependence of current response on different Cl position becomes less significant, as in the M1 case.

  \begin{table*}
  \caption{Total polarization magnitude ($|P|$), $z$ component ($|P_z|$), and band gap for MAPbI and MAPbICl structures (a, b, c and d shown in Fig.~\ref{cl_}) with both molecular orientations (M1 and M2).}\label{table_}
  \resizebox{\columnwidth}{!}{
  \begin{tabular}{ccccccccccccc}\hline\hline

&  \multicolumn{6}{c} {M1}   & \multicolumn{6}{c} {M2} \cr
Structure&  {\ } & $|P|$ & $|P_z|$ &{\ }& \multicolumn{2}{c}{Band gap (eV)} & {\ \ }  & $|P|$ &$|P_z|$&{\ }& \multicolumn{2}{c}{Band gap (eV)} \cr 
             &  {  } &  ($\mu$C/cm$^{2}$)&    ($\mu$C/cm$^{2}$){\ }&   & NSOC  &SOC & {\ \ }  &    ($\mu$C/cm$^{2}$)&  ($\mu$C/cm$^{2}$) {\ }&     &  NSOC & SOC \cr \hline
MAPbI &{\ \ }& 6.8 & 5.0 &{\ \ }& 1.72 & 0.69 &{\ \ \ }  & 0.7 & 0.5 &{\ \ }& 1.70 & 0.75 \cr
a     &{\ \ }& 8.3 & 7.2 &{\ \ }& 1.75 & 0.76 &{\ \ \ }  & 1.8 & 1.0 &{\ \ }& 1.72 & 0.80 \cr
b     &{\ \ }& 8.0 & 6.3 &{\ \ }& 1.90 & 0.83 &{\ \ \ }  & 2.6 & 2.4 &{\ \ }& 1.84 & 0.83 \cr
c     &{\ \ }& 6.9 & 6.2 &{\ \ }& 1.93 & 0.83 &{\ \ \ }  & 2.3 & 0.3 &{\ \ }& 1.93 & 0.83 \cr
d     &{\ \ }& 6.1 & 4.4 &{\ \ }& 1.96 & 0.84 &{\ \ \ }  & 3.3 & 3.2 &{\ \ }& 1.91 & 0.83 \cr\hline \hline
  \end{tabular}
  }
  \end{table*}
 
  In summary, we calculate the shift current responses and polarization magnitudes of MAPbI and MAPbICl from first principles with and without SOC. We find that the SOC does not substantially alter the spectrum, though it reduces the band gap. Rather, the MA orientation and Cl substitution position can strongly affect the shift current response. When the MA molecules' net dipole moments are aligned in parallel, the PbI$_3$ inorganic frame becomes more distorted, resulting in relatively large shift current responses. Conversely, when the molecules have opposite dipole moments, the structure is nearly symmetric, resulting in much smaller shift current responses. The substitution of Cl at the equatorial site can enhance the shift current response, because the orbital character contribution at the CBM is more delocalized along the shift current direction. Thus a higher shift current response can be obtained by introducing a large lattice distortion with MA molecules aligned in parallel, and by substituting Cl at equatorial positions.
  
 \begin{acknowledgement}
  F.Z. was supported by the Department of Energy Office of Basic Energy Sciences, under Grant Number DE-FG02-07ER46431. H.T. was supported by the Office of Naval Research, under Grant Number N00014-12-1-1033. F.W was supported by the Department of Energy Office of Basic Energy Sciences, under Grant Number DE-FG02-07ER46431. N.Z.K was supported by Office of Naval Research, under Grant Number N00014-14-1-0761, and by the Roy \& Diana Vagelos Scholars Program in the Molecular Life Sciences. A.M.R. was supported by the National Science Foundation, under Grant Number CMMI-1334241. Computational support was provided by the High-Performance Computing Modernization Office of the Department of Defense and the National Energy Research Scientific Computing Center.
 \end{acknowledgement}


\providecommand{\latin}[1]{#1}
\providecommand*\mcitethebibliography{\thebibliography}
\csname @ifundefined\endcsname{endmcitethebibliography}
  {\let\endmcitethebibliography\endthebibliography}{}


\begin{mcitethebibliography}{51}
\bibitem[dip()]{dipolemoment}
The dipole moment of molecule CH3NH3$^{+}$ is calculated as 2.2 Debye with
  HF/6-311G(p,d) using GAMESS.\relax
\mciteBstWouldAddEndPunctfalse
\mciteSetBstMidEndSepPunct{\mcitedefaultmidpunct}
{}{\mcitedefaultseppunct}\relax
\EndOfBibitem
\end{mcitethebibliography}
\end{document}